# Self-organized vanadium and nitrogen co-doped titania nanotube arrays with enhanced photocatalytic reduction of $CO_2$ into $CH_4$

Dandan Lu, Min Zhang[*], Zhihua Zhang, Qiuye Li, Xiaodong Wang and Jianjun Yang[*]

**Abstract**

Self-organized V-N co-doped $TiO_2$ nanotube arrays (TNAs) with various doping amount were synthesized by anodizing in association with hydrothermal treatment. Impacts of V-N co-doping on the morphologies, phase structures, and photoelectrochemical properties of the TNAs films were thoroughly investigated. The co-doped $TiO_2$ photocatalysts show remarkably enhanced photocatalytic activity for the $CO_2$ photoreduction to methane under ultraviolet illumination. The mechanism of the enhanced photocatalytic activity is discussed in detail.

## Background

Greenhouse gases such as $CO_2$ and chlorofluorocarbon (CFCs) are the primary causes of global warming. The atmospheric concentration of $CO_2$ has steadily increased owing to human activity, and this accelerates the greenhouse effect. The photocatalytic reduction of $CO_2$ is a promising technical solution since it uses readily available sunlight to convert $CO_2$ into valuable chemicals, such as methanol or methane, in a carbon friendly manner [1].

$TiO_2$ is a popular catalyst for photoreduction of $CO_2$ owing to the advantages of earth abundance, low toxicity, and chemical stability. Yet it has so far yielded only low carbon dioxide conversion rates despite using ultraviolet illumination for band gap excitations [2]. While the intrinsic idea of photocatalytic conversion of carbon dioxide and water (vapor) into hydrocarbon fuels is appealing, the process has historically suffered from low conversion rates. Numerous studies have been reported on how to increase the photoreduction activity of $TiO_2$ using transition metal-doped and/or modified $TiO_2$. Transition metal doping has been applied not only to modify the photoactivity of $TiO_2$ but also to influence the product selectivity. For example, mesoporous silica-supported Cu/$TiO_2$ nanocomposites showed significantly enhanced $CO_2$ photoreduction rates due to the synergistic combination of Cu deposition and high surface area $SiO_2$ support [3]. Dispersing Ce-$TiO_2$ nanoparticles on mesoporous SBA-15 support was reported to further enhance both CO and $CH_4$ production due to the modification of $TiO_2$ with Ce significantly stabilized the $TiO_2$ anatase phase and increased the specific surface area [4]. However, increasing the content of metal dopant does not always lead to better photocatalytic activity. The promotion of the recombination efficiency of the electron-hole pairs may be due to excessively doped transition metal.

Besides, nonmetal-doped $TiO_2$ have been used as visible light-responsive photocatalysts for $CO_2$ photoreduction. Significant enhancement of $CO_2$ photoreduction to CO had been reported for I-doped $TiO_2$ due to the extension of $TiO_2$ absorption spectra to the visible light region by I doping [5]. Enhanced visible light-responsive activity for $CO_2$ photoreduction was obtained over mesoporous N-doped $TiO_2$ with noble metal loading [6]. Nitrogen doping into $TiO_2$ matrix is more beneficial from the viewpoint of its comparable atomic size with oxygen, small ionization energy, metastable center formation and stability. However, a main drawback of N doping is that only relatively low concentrations of N dopants can be implanted in $TiO_2$.

In order to overcome the abovementioned limitations, modified $TiO_2$ by means of nonmetal and metal co-doping was investigated as an effective method to improve the photocatalytic activity. Among the current research of single ion doping into anatase $TiO_2$, N-doping and V-doping are noteworthy. Firstly, both elements are close neighbors of the elements they replace in the periodic table. They also share certain similar physical and chemical characteristics with the replaced elements. Secondly, impurity states of N dopants act as shallow acceptor levels, while those of V dopants act as shallow donor levels. This result in less recombination centers in the forbidden band of $TiO_2$ and thus prolongs the lifetime of photoexcited carriers [7]. So the co-doping of V and N into the $TiO_2$ lattice is of particular significance. Recently, V and N co-doped $TiO_2$ nanocatalysts showed enhanced photocatalytic activities for the degradation of methylene blue compared with mono-doped $TiO_2$ [8]. Wang et al. synthesized V-N co-doped $TiO_2$ nanocatalysts using a novel two-phase hydrothermal method applied in hazardous PCP-Na decomposition [9]. Theoretical and simulation work also found that N-V co-doping could broaden the absorption spectrum of anatase $TiO_2$ to the visible light region and increase its quantum efficiency [10]. However, the effect of V, N co-dopant in $TiO_2$ on the efficiency of $CO_2$ photocatalytic reduction has not been studied yet. In the present work, we made efforts to improve photocatalytic carbon dioxide conversion rates by the following strategies: (1) employ high surface area titania nanotube arrays, with vectorial charge transfer, and long-term stability to photo and chemical corrosion; and (2) modify the titania to enhance the separation of electron-hole pairs by incorporating nitrogen and vanadium. This article reports the synthesis, morphologies, phase structures, and photoelectrochemical of self-organized V, N co-doped $TiO_2$ nanotube arrays as well as the effect of V and N co-doping on photocatalytic reduction performance of $CO_2$ into $CH_4$.

* Correspondence: zm1012@henu.edu.cn; yangjianjun@henu.edu.cn
Key Laboratory for Special Functional Materials of Ministry of Education, Henan University, Kaifeng 475004, People's Republic of China

## Methods

### Fabrication of V, N co-doped $TiO_2$ nanotube arrays

V, N co-doped $TiO_2$ nanotube arrays (TNAs) were fabricated by a combination of electrochemical anodization and hydrothermal reaction. Firstly, highly ordered TNAs were fabricated on a Ti substrate in a mixed electrolyte solution of ethylene glycol containing $NH_4F$ and deionized water by a two-step electrochemical anodic oxidation process according to our previous reports [11]. Interstitial nitrogen species were formed in the TNAs due to the electrolyte containing $NH_4F$ [12]. Then, the amorphous TNAs were annealed at 500°C for 3 h with a heating rate of 10°C/min in air ambience to obtain crystalline phase. We denoted these single N-doped TNAs samples as N-$TiO_2$.

V, N co-doped TNAs were prepared by a hydrothermal process. As-prepared N-$TiO_2$ samples were immersed in Teflon-lined autoclaves (120 mL, Parr Instrument, Moline, IL, USA) containing approximately 60 mL of $NH_4VO_3$ aqueous solution (with different concentration 0.5, 1, 3, and 5 wt.%) as the source of both V and N. All samples were hydrothermally treated at 180°C for 5 h and then naturally cooled down to room temperature. Finally, all samples were rinsed with deionized water and dried under high purity $N_2$ stream. The corresponding samples (0.5%, 1%, 3%, and 5%) were labeled as VN0.5, VN1, VN3, and VN5. For control experiment, sample denoted as VN0 was prepared by the previously mentioned hydrothermal process in 60 mL pure water without $NH_4VO_3$ addition.

### Characterization

Surface morphologies of all samples were observed with field emission scanning electron microscope (FESEM, JEOL JSM-7001 F, Akishima-shi, Japan) at an accelerating voltage of 15 kV. Phase structures of the photocatalysts were analyzed by X-ray diffraction (XRD) analysis on an X'Pert Philips (Amsterdam, The Netherlands) diffractometer (Cu Kα radiation, 2θ range, 10° to 90°; step size, 0.08°). Chemical state and surface composition of the samples were obtained with an Axis Ultra X-ray photoelectron spectroscope (XPS, Kratos, Manchester, UK; a monochromatic Al source operating at 210 W with a pass energy of 20 eV and a step of 0.1 eV was used). All binding energies (BE) were referenced to the C 1 s peak at 284.8 eV of the surface adventitious carbon. UV-vis diffused reflectance spectra of N-$TiO_2$ and V, N-co-doped $TiO_2$ nanotube arrays were obtained using a UV-vis spectrophotometer (UV-2550, Shimadzu, Kyoto, Japan).

### Photoelectrochemical measurements

Photoelectrochemical experiments were monitored by an electrochemical workstation (IM6ex, Zahner, Germany). V, N co-doped TNAs (an active area of 4 $cm^2$) and platinum foil electrode were used as working electrode and counter electrode, and saturated calomel electrode (SCE) acted as reference electrode, respectively. 1 M KOH aqueous solution was used as the supporting electrolyte and purged with $N_2$ for 20 min before measurement to remove the dissolved oxygen. A 300-W Hg lamp was used as the light source. Photocurrent measurements were carried out under UV-vis irradiation at an applied bias voltage of 0.4 V (vs. SCE) in ambient conditions at room temperature.

### Photocatalytic reduction of $CO_2$

Photocatalytic reduction of $CO_2$ was performed in a 358-mL cylindrical glass vessel containing 20 mL 0.1 mol/L $KHCO_3$ solution with a 300-W Hg lamp fixed parallel to the glass reactor as light source. TNAs films were placed in the center of the reactor before sealing the reactor. Prior to reduction experiment under irradiation, ultra-pure gaseous $CO_2$ and water vapor were flowed through the reactor for 2 h to reach adsorption equilibrium within the reactor. Each experiment was followed for 6 h. The analysis of $CH_4$ was online conducted with a gas chromatography (GC).

## Results and discussion
### Morphology

Figure 1 shows FESEM images of N-TiO$_2$ and V, N co-doped TNAs with various doping amounts. N-TiO$_2$ nanotube arrays before hydrothermal treatment are uniformly stacked with tubular structures with an average diameter of 130 nm and an average wall thickness of 20 nm (Figure 1a). The side view image in Figure 1b also reveals that the vertically orientated nanotubes have an average length of 11 μm. According to SEM observations in Figure 1c,d, the VN0 sample after hydrothermal treatment in pure water presents no apparent structural transformation. The side view image in Figure 1d also shows the highly ordered nanotube arrays with similar diameter and wall thickness of N-TiO$_2$ sample before hydrothermal reaction. Yu et al. had reported that the nanotube array structures were completely destroyed after 180°C hydrothermal treatments with TNAs samples due to the enhanced anatase crystallinity and phase transformation from amorphous to anatase [13]. In our experiments, oxidized TNAs samples were calcinated at 500°C to realize phase transformation from amorphous to anatase before hydrothermal process. By this way, the reported hydrothermally induced collapse was prevented with a simple calcination step. All hydrothermal-treated TNAs samples including the V, N co-doped TNAs show no apparent morphology change after hydrothermal co-doping process. Figure 1e,f presents the top and side view images of the V, N co-doped TNAs with maximal doping amounts of 5% in our experiments. Nanotube arrays structure is kept unchanged in VN5 sample after hydrothermal co-doping process as shown in Figure 1f. We found that appropriate doping amount of V and N does not change the diameter of nanotubes. However, excessive dopants may lead to some particles or aggregates on the surface of nanotube arrays and some even block the pores and channels.

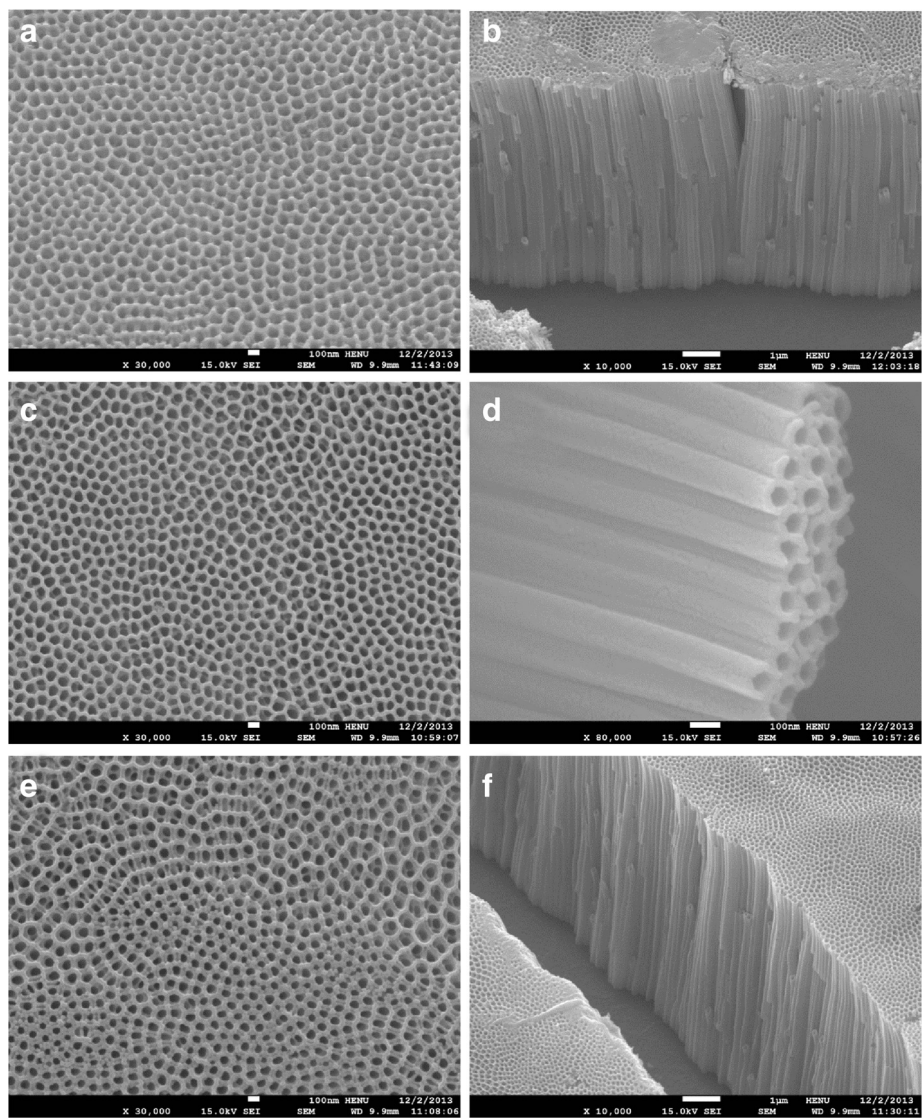

**Figure 1** FESEM top views and side views for N-TiO$_2$ (a, b), VN0 (c, d), and VN5 (e, f).

### Crystal structure

The structural analysis of doped $TiO_2$ nanotube arrays was usually carried out by using X-ray diffraction (XRD) and Raman spectroscopy [14]. Here, XRD measurements were performed to investigate the changes of phase structures of N-$TiO_2$ sample and V, N co-doped TNAs with various doping amounts. As shown in Figure 2, diffraction peaks of all samples were ascribed to pure anatase $TiO_2$ diffraction pattern consistent with the values in the standard card (JCPDS card no. 21-1272) [15]. No significant characteristic peak of vanadium species is found in corresponding XRD patterns. Numerous reports showed that the incorporation of transition metal ions into other compounds as dopant could distort the original crystal lattice of the doped materials [16]. A detail analysis of XRD patterns was performed by enlarging the anatase (101) plane of the samples as shown in the inset of Figure 2. Compared with N-$TiO_2$, the peak position of the V, N co-doped TNA samples gradually shifted toward a higher diffraction angle. It suggests that the V ions might be successfully incorporated into the crystal lattice of anatase $TiO_2$ as vanadyl groups ($V^{4+}$) or polymeric vanadates ($V^{5+}$) and substituted for $Ti^{4+}$ because the ionic radii of $V^{4+}$ (0.72 Å) and $V^{5+}$ (0.68 Å) were both slightly smaller than that of $Ti^{4+}$ (0.75 Å) [17]. However, peak position change of VN5 was not obvious, indicating that the doped V ions might be excessive and aggregate on the surface of TNAs and then inhibit the incorporation of ions into crystal lattice. For VN0 sample without co-doping, its crystal lattice did not change through the hydrothermal process and kept the similar peak position with N-$TiO_2$ sample.

### XPS analysis

Figure 3 shows the high-resolution XPS spectra of Ti 2p, O 1 s, N 1 s, and V 2p regions for N-$TiO_2$, VN0, and VN3 samples. A significant negative shift is found for Ti 2p in Figure 3a and O 1 s in Figure 3b when V and N were co-doped into $TiO_2$ by hydrothermal process. The measured binding energies of Ti $2p_{3/2}$ and O 1 s for N-$TiO_2$ and VN0 are 458.7 and 529.9 eV, respectively. As compared to N-$TiO_2$ and VN0, the binding energy of Ti $2p_{3/2}$ for the VN3 sample is shifted to 458.5 eV. The lower binding energy of Ti 2p in co-doped $TiO_2$ suggests the different electronic interactions of Ti with ions and substitutes for Ti [9], which further justifies the incorporation of vanadium and nitrogen into the $TiO_2$ lattice. The binding energy of O 1 s of $TiO_2$ lattice oxygen (Ti-O-Ti) for the VN3 sample is also shifted to 529.6 eV. Oxygen molecules can be dissociatively absorbed on the oxygen vacancies induced by doping N, thereby leading to a slight shift to lower binding energy of O 1 s of $TiO_2$ lattice oxygen (Ti-O-Ti) [18].

Figure 3c shows the high-resolution XPS spectra and corresponding fitted XPS for the N 1 s region of N-$TiO_2$, VN0, and VN3. A broad peak extending from 397 to 403 eV is observed for all samples. The center of the N 1 s peak locates at ca. 399.7, 399.6, and 399.4 eV for N-$TiO_2$, VN0, and VN3 samples, respectively. These three peaks are higher than that of typical binding energy of N 1 s (396.9 eV) in TiN [19], indicating that the N atoms in all samples interact strongly with O atoms [20]. The binding energies of 399.7, 399.6, and 399.4 eV here are attributed to the oxidized nitrogen similar to $NO_x$ species, which means Ti-N-O linkage possibly formed on the surface of N-$TiO_2$, VN0, and VN3 samples [21-23]. The concentrations of V and N in VN3 derived from XPS analysis were 3.38% and 4.21% (at.%), respectively. The molar ratios of N/Ti on the surface of N-$TiO_2$ and VN3 were 2.89% and 14.04%, respectively, indicating obvious increase of N doping content by hydrothermal treatment with ammonium metavanadate. As shown in Figure 3d, the peaks appearing at 516.3, 516.9, 523.8, and 524.4 eV could be attributed to $2p_{3/2}$ of $V^{4+}$, $2p_{3/2}$ of $V^{5+}$, $2p_{1/2}$ of $V^{4+}$, and $2p_{1/2}$ of $V^{5+}$ [24,25]. It was established that the $V^{4+}$ and $V^{5+}$ ions were successfully incorporated into the crystal lattice of anatase TiO$_2$ and substituted for Ti$^{4+}$ ions.

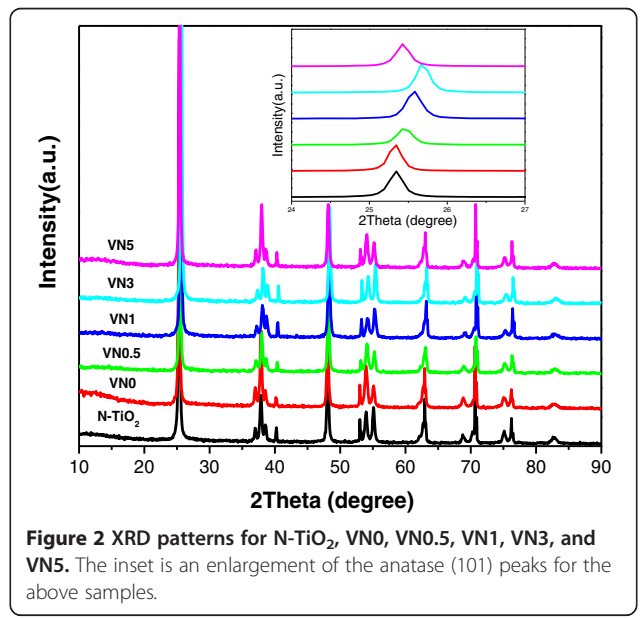

**Figure 2** XRD patterns for N-$TiO_2$, VN0, VN0.5, VN1, VN3, and VN5. The inset is an enlargement of the anatase (101) peaks for the above samples.

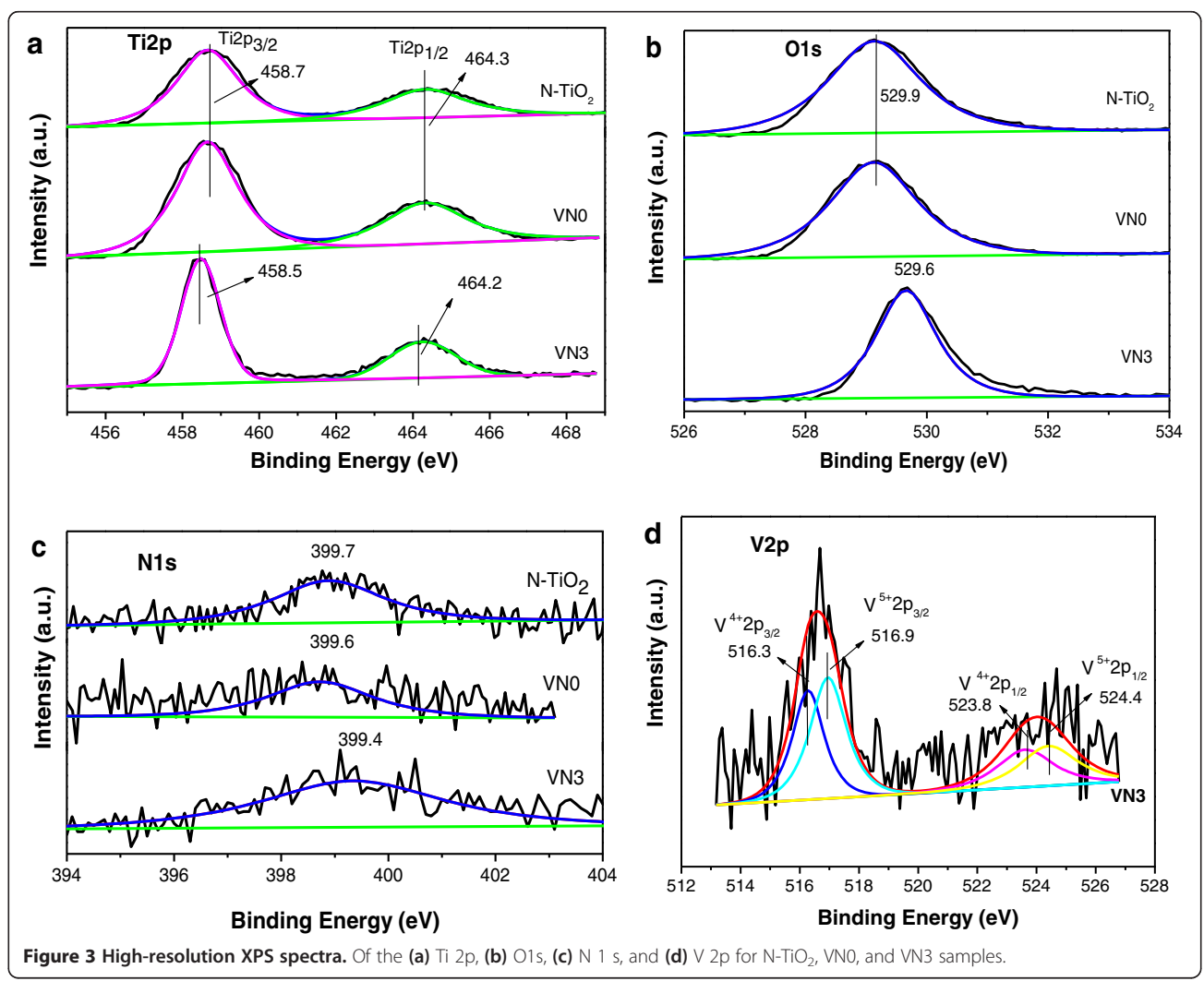

**Figure 3 High-resolution XPS spectra.** Of the (a) Ti 2p, (b) O1s, (c) N 1 s, and (d) V 2p for N-TiO$_2$, VN0, and VN3 samples.

### UV-vis DRS spectra analysis

UV-vis diffuse reflectance spectra of N-TiO$_2$ and V, N co-doped TiO$_2$ nanotube arrays are displayed in Figure 4. The spectrum obtained from the N-TiO$_2$ shows that N-TiO$_2$ primarily absorbs the ultraviolet light with a wavelength below 400 nm. For the V, N co-doped TNAs samples of VN0.5 and VN1, the UV-vis diffuse reflectance spectroscopy (DRS) spectra present a small red shift of adsorption edge and a higher visible light absorbance. With the increase of co-doping amount, an obvious red shift of the absorption edge and enhanced visible light absorbance were observed for the VN3 and VN5 samples. However, no obvious change of visible light absorbance was found for VN0, which indicates that the visible light absorbance of co-doped samples may be due to the contribution of both interstitially doped N and substitutionally doped V. Kubelka-Munk function was used to estimate the band gap energy of all samples by plotting $(\alpha h\nu)^{1/2}$ vs. energy of absorbed light. The calculated results as shown in Figure 4b indicated that the band gap energies for N-TiO$_2$, VN0, VN0.5, VN1, VN3, and VN5 are 3.15, 3.15, 2.96, 2.92, 2.42, and 2.26 eV, respectively. It shows that the V, N co-doped TiO$_2$ nanotube array samples have a narrower band gap than that of N-TiO$_2$. When the dopant exists in an optimal doping concentration, V$^{4+}$ and V$^{5+}$ can easily substitute Ti$^{4+}$ in the TiO$_2$ lattice, which might produce the new energy level and extend the light adsorption of TiO$_2$ due to the fact that ionic radius of V$^{4+}$ and V$^{5+}$ were both slightly smaller than that of Ti$^{4+}$ [17]. Overall, the UV-vis DRS results indicate that N and V co-doped TiO$_2$ nanotube arrays are more sensitive to the visible light than N-TiO$_2$ samples.

### Photoelectrochemical properties

A series of the photoelectrochemical (PEC) experiments were carried out to investigate the effect of the V, N co-doping of TNAs films on the charge carriers separation and electron transfer processes. As shown in Figure 5, prompt generation of photocurrents was observed for all TNA samples upon illumination at an applied potential of 0.4 V vs. SCE. All samples showed good photoresponses and highly reproducible for numerous on-off cycles under the light on and light off conditions. The V, N co-doped TNAs exhibited higher photocurrents than that of N-TiO$_2$ samples under UV irradiation. Herein, N-TiO$_2$ electrode shows that only a lower photocurrent density of 2.5 mA/cm$^2$ may be due to the rapid recombination of charge carriers. With the co-doping of V and N, the VN3 sample exhibited highest photocurrent (5.0 mA/cm$^2$) with optimal concentration. These results further inferred that V, N co-doped TiO$_2$ nanotube arrays possess good photoresponsivity to generate and separate photo-induced electrons and holes [26]. Excessive vanadium and nitrogen content caused the detrimental effect, which acted as recombination centers to trap the charge carriers and resulted in low quantum yields [2,27]. From the PEC experimental results, optimum content of V and N co-doped into TiO$_2$ play an important role in maximizing the photocurrent density mainly attributed to the effective charge carrier separation and improve the charge carrier transportation.

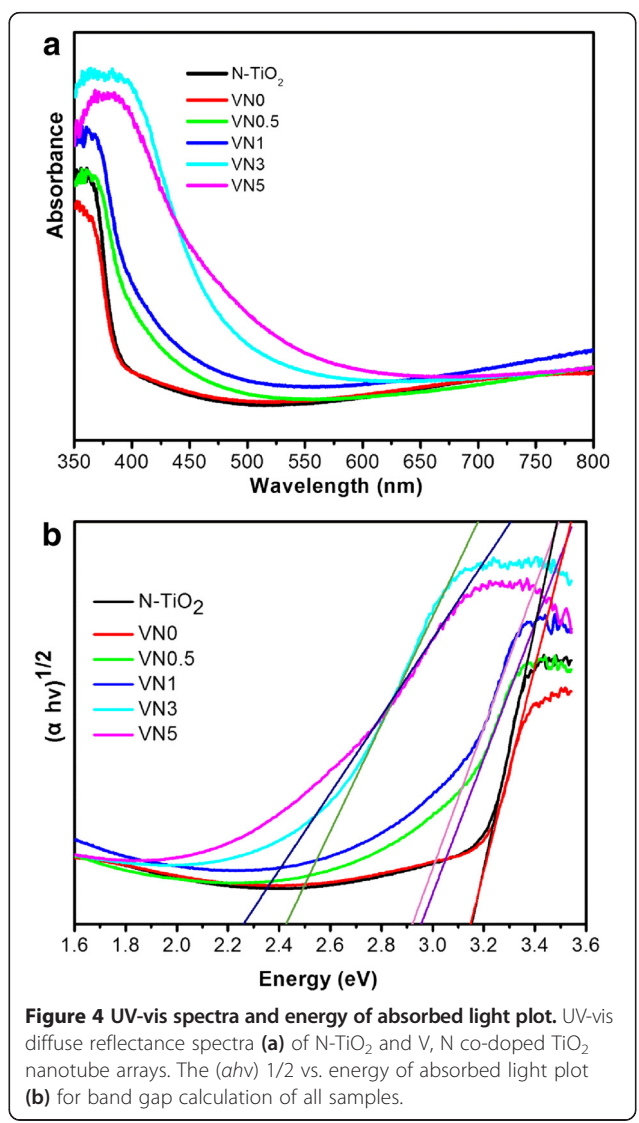

**Figure 4 UV-vis spectra and energy of absorbed light plot.** UV-vis diffuse reflectance spectra **(a)** of N-TiO$_2$ and V, N co-doped TiO$_2$ nanotube arrays. The ($\alpha h\nu$) 1/2 vs. energy of absorbed light plot **(b)** for band gap calculation of all samples.

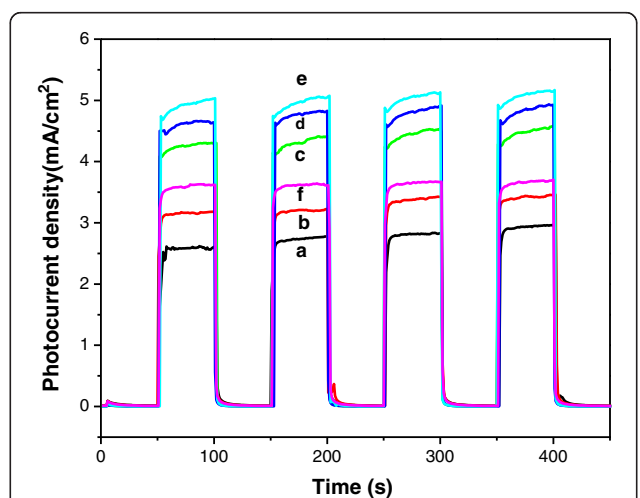

**Figure 5 Photocurrent responses in light on-off process at applied voltage.** Of 0.4 V (vs. SCE) under UV irradiation for (curve a) N-TiO$_2$, (curve b) VN0, (curve c) VN0.5, (curve d) VN1, (curve e) VN3, and (curve f) VN5.

### Photocatalytic reduction performance

Photoreduction of CO$_2$ to methane were performed as a probe reaction to evaluate the photocatalytic activity of the V, N co-doped TNA films. During the CO$_2$ photoreduction reaction, the increase of CH$_4$ concentration (ppm/cm$^2$, $\triangle$CH$_4$, which is the difference between CH$_4$ concentration at $t$ reaction time and the initial time) was used to evaluate the photocatalytic performance. As shown in Figure 6, concentration of CH$_4$ increased almost linearly with the UV irradiation time for the photocatalyst. V and N co-doped TNAs possess much higher photocatalytic activity than N-TiO$_2$ and VN0 sample hydrothermal treated in pure water. Moreover, photoreduction activity of V, N co-doped TNAs was enhanced and then decreased with the increase of doping content of vanadium and nitrogen. VN3 sample had the highest methane yield of 64.5 ppm h$^{-1}$ cm$^{-2}$. For comparison, reference reactions without catalysts or light irradiation were performed with other conditions being kept unchanged. All results indicated that there was almost no methane production when the experiment was carried out in the absence of catalysts or irradiation. We also investigated the effect of hydrothermal treatment on the photocatalytic activity. VN0 sample was obtained by the hydrothermal treatment of N-TiO$_2$ in pure water and used as a photocatalyst. A slightly enhanced photocatalytic activity was found for VN0 sample as shown in Figure 6. This hydrothermal-assisted photocatalytic enhancement results are also confirmed by some researchers [28,29]. All results indicate that photo-excited process of V, N co-doped TNAs is essential in photoreduction process of $CO_2$. However, for VN5 sample, the reduction activity is the lowest one because a further increase in the vanadium content would result in the aggregation of dopant nanoparticles, fast recombination of hole and electron pairs, and excess oxygen vacancies and $Ti^{3+}$ defects state induced by nitrogen doping also served as recombination centers [30].

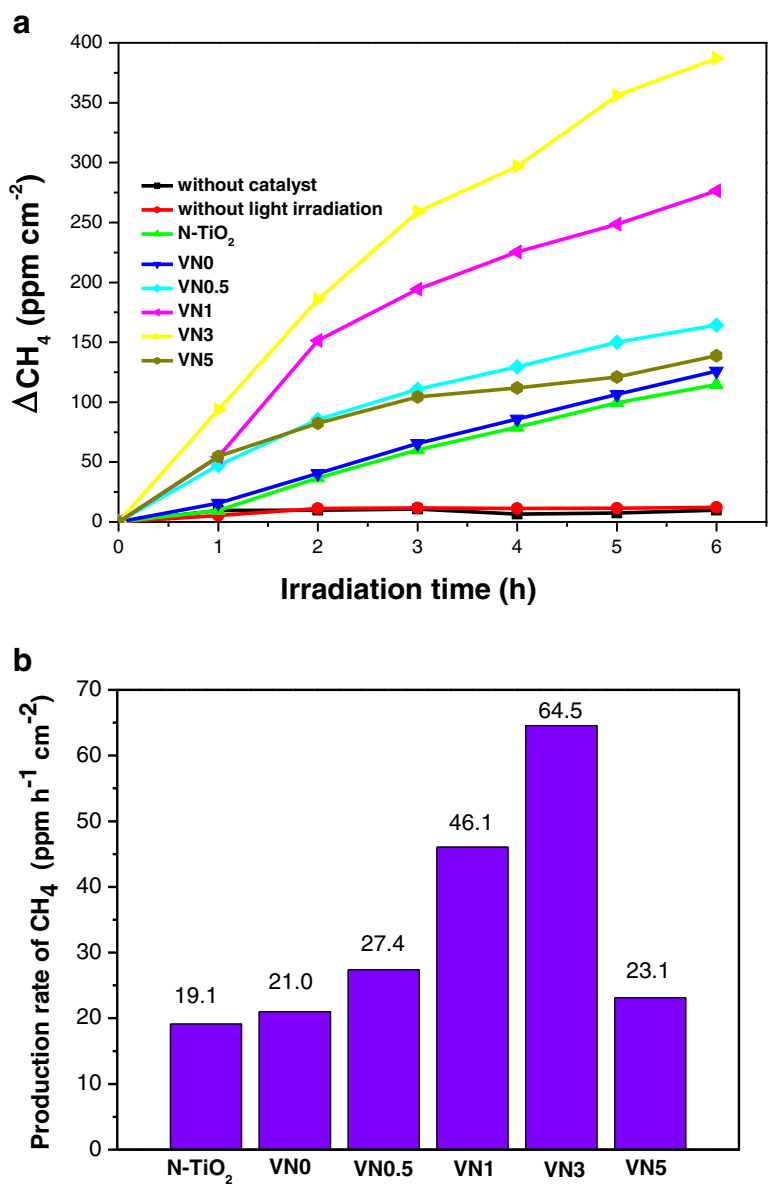

**Figure 6** △CH$_4$ concentration dependence on irradiation time (a) and production rate of CH$_4$ (b) for all catalysts under UV irradiation.

The photoreduction reaction of $CO_2$ over VN3 sample was also repeated to check the durability of photocatalyst. Figure 7 shows the CH$_4$ formation by VN3 sample for three times. After each cycle (6 h irradiation), the reaction vessel was degassed, then $CO_2$ and water vapor was introduced into it again. The photocatalytic activity could be restored after three cycles. In each cycle, the initial CH$_4$ evolution rate was recovered, and there was no CH$_4$ formation evolved when the light was off. The above durability results indicate that the V, N co-doped TNAs were stable under the present experimental conditions during the long irradiation time.

### Photocatalytic reduction mechanism

When TNAs were radiated by the light with photon energy higher or equal to the band gaps of $TiO_2$, more electrons and holes induced by V and N co-doping lead to the reduction of $CO_2$ successfully. Previous studies revealed the trapping of the excited electron and hole by oxygen vacancy and doped nitrogen respectively reduced the recombination rate. The presence of nitrogen dopants was considered to reduce the formation energy of oxygen vacancies [31]. At the same time, the existence of O vacancies stabilized the N impurities [32]. When oxygen vacancies and N impurities are incorporated together, there is an electron transfer from the higher energy 3d band of $Ti^{3+}$ to the lower energy 2p band of the nitrogen impurities [33]. The efficient separation and transfer of election-hole pairs might also be associated with the interaction of $V^{4+}$ and $V^{5+}$. The $V^{5+}$ species reacted with the electrons to yield $V^{4+}$ species, which on surface oxygen molecules generated the oxidant superoxide radical ion $O_2^-$. $O_2^-$ reacted with $H^+$ to produce hydroxyl radical and $H^+$ and $CO_2$ trapped electrons to produce $^\cdot H$ and $^\cdot CO_2^-$, which further reacted with holes to yield the final product, methane [34].

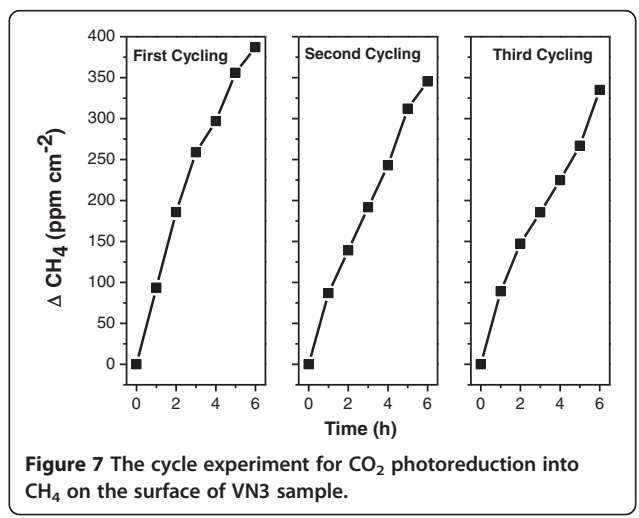

Figure 7 The cycle experiment for $CO_2$ photoreduction into $CH_4$ on the surface of VN3 sample.

$$TiO_2 \rightarrow h^+ + e^- \text{ (UV irradiation)}$$

$$V^{5+} + e^- \rightarrow V^{4+} + O_2 \rightarrow V^{5+} + O_2^-$$

$$O_2^- + 2H^+ \rightarrow {^\cdot}OH + {^\cdot}OOH$$

$${^\cdot}OH + H_2O + 3h^+ \rightarrow 3H^+ + O_2$$

$$H^+ + e^- \rightarrow {^\cdot}H$$

$$CO_2 + e^- \rightarrow {^\cdot}CO_2^-$$

$${^\cdot}CO_2^- + 8{^\cdot}H + h^+ \rightarrow CH_4 + H_2O$$

Superabundant V and N could result in a decrease of photoreduction activity for increasing recombination centers of electrons and holes.

## Conclusions

V-N co-doped $TiO_2$ nanotube arrays have been fabricated by a simple two-step method. V and N co-doped $TiO_2$ photocatalysts exhibit fine tubular structures after hydrothermal co-doping process. XPS data reveal that N is found in the forms of Ti-N-O and V incorporates into the $TiO_2$ lattice in V-N co-doped TNAs. V and N co-doping result in remarkably enhanced activity for $CO_2$ photoreduction to $CH_4$ due to the effective separation of electron-hole pairs. Meanwhile, the unique structure of co-doped $TiO_2$ nanotube arrays promoted the electron transfer and the substance diffusion.

**Competing interests**
The authors declare that they have no competing interests.

**Authors' contributions**
DDL carried out the synthesis, characterization, and photocatalytic reduction experiments. ZHZ participated in the synthesis and SEM characterization experiments. QYL and XDW participated in the XPS and Raman characterizations. MZ and JJY participated in the design and preparation of the manuscript. All authors read and approved the final manuscript.

**Acknowledgements**
The authors thank the National Natural Science Foundation of China (no.21203054) and Program for Changjiang Scholars and Innovation Research Team in University (no. PCS IRT1126).